\documentclass[12pt,english]{article}
\usepackage{graphicx}
\usepackage{cite}
\usepackage{babel}
\newcommand{\br}{\mbox{\boldmath $r$}}
\newcommand{\bp}{\mbox{\boldmath $p$}}
\newcommand{\bL}{\mbox{\boldmath $L$}}
\newcommand{\bS}{\mbox{\boldmath $S$}}

\begin{document}
\title{SPIN-ORBIT PENDULUM IN DIRAC OSCILLATOR
\thanks{Research partially supported by KBN grant No.~5 P03B 010 20 and
Technical University of Zielona G\'ora}}
\author{{\Large Marcin Turek }\\
 Institute of Physics, Maria Curie-Sk\l odowska University,\\
  20-031 Lublin, Poland\\
  \and {\Large Piotr Rozmej}\\
Division of Mathematical Methods in Physics, Institute of Mathematics,\\ 
Technical University of Zielona G\'ora, 65-246 Zielona G\'ora, Poland
}

\maketitle

\begin{abstract}
The dynamics of wavepackets in a relativistic Dirac oscillator (DO) is
considered. A comparison to nonrelativistic {\em spin-orbit pendulum} effect 
is discussed. Particular relativistic effects, like {\em Zitterbewegung} in
spin motion, are found in Dirac representation. This trembling motion  
disappears in Foldy-Wouthuysen representation. A substantial difference 
between the dynamics of wavepackets corresponding to circular and linear 
orbits of a particle is obtained and discussed.
\end{abstract}

\section{Introduction}

Few years ago we were considering \cite{arv1,arv2,roz1,roz2} a 
motion of a wave packet (WP) representing a fermion in a
spherical symmetric potential (e.g. the nonrelativistic three-dimensional
 harmonic oscillator -- 3dHO) with a spin orbit coupling.
 We have discovered a phenomenon called {\em spin-orbit pendulum}, 
 in which the spin of the particle, 
 initially polarized along a well-defined direction makes a
 reversible transition to a state in which no direction is preferred.  
In other words, one can observe a reversible transfer of the angular momentum
between the spin and orbital motion. This behavior is very similar to revivals
of population inversions within the Jaynes-Cummings model.

We would like to present the relativistic version of the spin-orbit pendulum 
using the Dirac oscillator \cite{ito,cook,ui,cho,mosh}
instead of the 3dHO. The mechanism of spin collapses and
revivals remains the same, but we have found several new effects caused by
the relativistic regime: broken periodicity, the {\em Zitterbewegung} 
of spin mean values and the appearence of the WP components, 
that are made of negative energy states
and can be interpreted within the hole theory.

In contrast to the work \cite{do}, where the WP motion corresponding to 
circular orbits has been presented for DO, we discuss here initial
conditions leading to a different topology of the orbital motion,
{\em i.e.} linear orbits and then compare results with that of circular ones. 
Results presented in  \cite{do} and in the present paper 
are the first ones on 3d wavepacket dynamics in DO. The results presented
in \cite{nog,nog1} correspond to the relativistic 1+1 dimensional case 
of Dirac oscillator.

\section{The Dirac Oscillator}

A system called by Moshinsky and Szczepaniak \cite{mosh,moshinsky}
the Dirac oscillator (DO) had been considered several times in the past 
 \cite{ito,cook,ui,cho} in different contexts. Therefore here we only recall
 the most important relations relevant for a construction of WP and its 
 evolution. The DO is described by the equation:
\begin{equation}\label{DO}
i \hbar { { \partial \Psi} \over { \partial t} } = 
H_{DO} \Psi= c \left[ \, \mbox{\boldmath $\alpha $} \cdot (\, 
\bp- i m \omega \br \beta ) + m c \beta \right] \, ,
\end{equation}
where $ \mbox{\boldmath $ \alpha $}, \beta $ are usual Dirac matrices.
 It was shown that both the large and small components of an DO eigenstate obey
 equations:
\begin{equation} 
(E^2-m^2 c^4) \, \Psi_{l(s)}=\Big[ \, c^2 (\bp^{2} +m \omega^2 \br^{2} ) \mp 
3 \hbar\omega m c^2\,\mp { {4 m c^2 \omega} \over {\hbar} } \bL \cdot \bS
 \, \,  \Big] \, \Psi_{l(s)} \, .
\end{equation} 
Expression on the right side is, except an irrelevant constant, a
three-dimensional harmonic oscillator (3dHO) Hamiltonian with a very strong
spin-orbit coupling term. Hence, both the components must be of the form
$ | N l j m_j \rangle  $.
The energy spectrum is given by
\begin{equation} \label{spectr}
E_{Nlj}= \pm \, mc^2 \sqrt{1+rA_{Nlj}}, \hbox{ \ \ \ \ \ \ \ \ \ }
r={{\hbar \omega} \over {m c^2}}, 
\end{equation}
where
\begin{equation}
A_{Nlj}= \left\{ \matrix{ 2(N-j)+1 & \hbox{ \ \ \ for } j = l+{{1} \over {2}} \cr
 2(N+j)+3 & \hbox{ \ \ \ for } j = l-{{1} \over {2}} \cr 
}
\right.
\end{equation}
The parameter $ r $ enables, if it is small enough, a transition to the
nonrelativistic limit -- quantity $ E-mc^2 $ gets nearly equal to the energy 
spectrum of the nonrelativistic 3dHO with LS coupling.
 A requirement that $ \Psi_l $ and $ \Psi_s $ belong 
to the same values of $ E $ and $ j $ results in the following form
of normalized eigenstates corresponding to positive and negative energy
 eigenvalues, respectively
\begin{eqnarray} \label{pos}
 \Psi_+(t) &=& \left( \, \matrix{ \sqrt{ {1\over 2} \, \left(1 + { { \omega_0}
 \over { \omega_{Nlj} } }   \right) }  \, | N l j m_j \rangle \cr
sgn  \sqrt{ {1\over 2} \, \left( 1 - { { \omega_0} \over { \omega_{Nlj} } }
\right)  }
 \, | N-1 l' j m_j \rangle
}  \, \right) \,  \exp { ( - i \omega_{Nlj}  \, ) }\, , \\ \label{neg}
  \Psi_-(t) &=& \left( \, \matrix{ \sqrt{ {1\over 2} \, \left(1 - { { \omega_0}
 \over { \omega_{Nlj} } }   \right) }  \, | N l j m_j \rangle \cr
sgn  \sqrt{ {1\over 2} \, \left( 1 + { { \omega_0} \over { \omega_{Nlj} } }
\right)  }
 \, | N-1 l' j m_j \rangle
}  \, \right) \, \exp { (\, i \omega_{Nlj} \, ) } \, .
\end{eqnarray} 
The following notations have been introduced:
\begin{equation}
 l'= \left\{  \matrix{ \, l+1 \cr
 \, l-1  \cr 
} \right. \, ,
\hbox{ \ \ \  and \ \ \  }
 sgn = \left\{  \matrix{ \, -i & \hbox{ \ \ \ \ \ \ \ for \ } j=l+{1 \over 2} \cr
\,  i & \hbox{ \ \ \ \ \ \ \  for \ } j=l-{1 \over 2}  \cr 
} \right.  \, ,
\end{equation}
$ \omega_0= (mc^2)/\hbar$ 
and $\omega_{Nlj} =| E_{Nlj}| \slash \hbar$.
An exact Foldy-Wouthuysen (FW) tranformation \cite{FW}
for DO is also known \cite{mor}. In this
representation states with negative energy disappear,
which makes calculations much easier.
The spectrum is the same as in the case of Dirac representation~(\ref{spectr}).

\section{Initial form of the WP}

We study the evolution of a Gaussian-shape WP which is initially    
centered at $\mathbf r_0 $ and has the average momentum $ \mathbf p_0$ (i.e.
 a 3dHO coherent state). Moreover, the initial WP is an eigenstate of the
 spin pointed at some arbitrary direction defined by $\alpha$ and $\beta $, 
 which without loosing of the generality could be chosen as real numbers
 \begin{equation}
 \label{gauss}
 \Psi({\mathbf r}, t=0)= {{1} \over {(2 \pi)^{ 3 \over 4} \sigma^{3 \over 2} 
 }} \exp{ \left[  {{({{\mathbf r} -{ \mathbf r_0}})^2} \over {2 \sigma^2}} + i 
 {{{{\mathbf p_0} \cdot { \mathbf r}}} \over {\hbar}}
 \right] } \, \pmatrix{ \alpha \cr \beta \cr 0 \cr 0} \, ,
 \end{equation}
 where $ \sigma= \sqrt{\hbar \slash m \omega}$. 
 
 In the following we will present the behavior of two kinds of the WP
 (\ref{gauss})
 corresponding to the particular choices of initial conditions:
 \begin{itemize}
 \item The {\em circular} WP which is obtained by choosing
\begin{equation}
{\mathbf r_0} = x_0 \, \hat x = \sigma \sqrt{N} \, \hat x \hbox{ \ \ \ \ \ }
{\mathbf p_0} = p_0 \, \hat y = \sigma^{-1}\sqrt{N} \, \hat y \, .
\end{equation}  
Such WPs have already been considered in our previous
papers for nonrelativistic HO \cite{arv1,roz2} and relativistic DO \cite{do}.
\item The {\em linear} WP defined by the following choise
\begin{equation} \label{lin}
{\mathbf r_0} = z_0 \, \hat z \hbox{ \ \ \ \ \ }
{\mathbf p_0} = p_0 \, \hat z  \; .
\end{equation}
Motion of linear WPs for nonrelativistic HO have been discussed in 
\cite{ber,ber1}.
 \end{itemize}
Symbols $\hat x, \hat y, \hat z$ denote unit vectors along the relevant axis.  
The words {\em circular} and { \em linear} are used to emphasize different
shapes of trajectories of considered coherent states.
In a pure HO case such WPs move along circular and linear classical
trajectories, respectively, without spreading.  

\section{Scheme of calculations}
 
The circular WP at $ t=0 $ could be written in the basis of 3dHO
eigenstates;
\begin{equation} \label{cir} 
\Psi_{\rm cir}=\sum_{l=0}^{\infty} \lambda_{l} \, 
| l,l,l \rangle \pmatrix{ \alpha \cr \beta \cr 0 \cr 0} \, ,
\end{equation}
where decomposition coeficients are given by:
\begin{equation}
 \lambda_l = (-1)^l \exp{(- N \slash 2}) \, {{{N}^{l \slash 2}} \over
 {\sqrt{l!}}} \, .
 \end{equation}
Note, that this expression involves states with $ N=l=m $ only,
taken with Poissonian weights. 
The initial form of the linear WP is more complex
\begin{equation} \label{lin1} 
\Psi_{\rm lin}=\sum_N^{\infty}\sum_{l=0(1)}^{N} \lambda_{Nl} \, 
| N,l,0 \rangle \pmatrix{ \alpha \cr \beta \cr 0 \cr 0} \, .
\end{equation}
 Introducing a complex parameter 
 \begin{equation}
\zeta ={1 \over \sqrt{2} } \, \left(  \sqrt{ {{m \omega} \over \hbar }} 
 \, z_0 + i \, \sqrt{ \hbar \over{ m \omega  }} \, p_0  \right)   
 \end{equation}
 we can write these coefficients as
 \begin{equation}
 \lambda_{Nl}=(-1)^{j} \, \sqrt{{l! \, (2l+1)!! \, 2^{l-j}}
 \over{(2l)!\, j!\,(N+l+1)!!}} \, \exp{\left( -{ |\zeta|^2 }\over { 2}\right)}
\, \zeta^{N},  \hbox{ \ \ \ where \ \ \  } j={{N-l} \over {2}}  \,.
 \end{equation}
 Due to cylindrical symmetry with respect to the $Oz$ axis the WP (\ref{lin1}) 
 contains states with $ m=0 $ only.
 Next step is to obtain an explicit form of the WP at an arbitrary time instant.
It is convenient to perform a transformation to the coupled basis $ |Nljm_j
\rangle $ because using (\ref{pos}) and (\ref{neg}) we easily get a simple
formula describing the evolution of the single $ |Nljm_j
\rangle $ state:
\begin{equation}
\left( \matrix{  | N l j m_j \rangle \cr 0  } \right)_t =
 \left( \matrix{ A_{Nlj}(t) \, \, | N l j m_j \rangle \cr
 B_{Nlj}(t) \, \, | N-1 l' j m_j \rangle } \right) \,  ,
 \end{equation}
 where time-dependent coefficients $A$ and $ B $ are defined as
 \begin{eqnarray}
 A_{Nlj}(t) &=&   \cos ( \omega_{Nlj} t ) - i \, { {\omega_0} \over
 {\omega_{Nlj}}} \, \sin ( \omega_{Nlj} t ) 
\\
 B_{Nlj}(t) &=& -i \, sgn \,  \sqrt{ {1 \over 2} \, 
  \left( 1 - {{ \omega^2_0} \over { \omega_{Nlj}^2} } \right) } \,  
  \sin ( \omega_{Nlj} t ) 
  \, . 
  \end{eqnarray}
These formulae allow us to obtain the form of the WP at given time $ t $.
However, for further calculations, {\em e.g.} evaluation of the spin averages, 
we return to the decoupled basis $ | Nlm \rangle |s m_s \rangle $,
which is more useful for this purpose.
As a result of this procedure we obtain
\begin{equation}
| \Psi_{lin} (t)\rangle = \left( \, \matrix{
 | \psi_1 \rangle_t \cr | \psi_2 \rangle_t  \cr
  | \psi_3 \rangle_t \cr | \psi_4 \rangle_t  \cr
} \, \right)  \, .
\end{equation}
Every component of this bispinor has rather complicated form
\begin{eqnarray}
| \psi_1 \rangle_t &=& \sum_{Nl} \lambda_{Nl} \, \left\{
\,  \alpha \, \Big( \, {{l+1} \over {2l+1} }\, A_{Nl>}(t)  +
 {{l} \over {2l+1} } \, A_{Nl<}(t) \, \Big) \, | N,l,0 \rangle \, + 
  \right. \nonumber   \\  
&+& \left.  \beta \, { {\sqrt{l(l+1)}} \over {2l+1} } \, \Big( 
\,  A_{Nl>}(t)-A_{Nl<}(t) 
\, \Big) \, | N,l,-1 \rangle \,    \right\} \, , 
\\
| \psi_2 \rangle_t &=& \sum_{Nl} \lambda_{Nl} \, \left\{
\,  \beta \, \Big( \, {{l+1} \over {2l+1} }\, A_{Nl>}(t)  +
 {{l} \over {2l+1} } \, A_{Nl<}(t) \, \Big) \, | N,l,0 \rangle 
\, \right. + \nonumber    \\
  &+& \left. \alpha \, { {\sqrt{l(l+1)}} \over {2l+1} } \,
 \Big( \,  A_{Nl>}(t)-A_{Nl<}(t) 
\, \Big) \, | N,l,1 \rangle \,    \right\} \, ,
\\
| \psi_3 \rangle_t &=& \sum_{Nl} \lambda_{Nl} \, \left\{ \,
-  \alpha \,{{l+1} \over {\sqrt{(2l+1)(2l+3)}} } \, B_{Nl>}(t) \, 
| N-1 , l+1, 0\rangle \,+ \right. \nonumber  \\
&-&  \alpha \,{{l} \over {\sqrt{(2l+1)(2l-1}} } \, B_{Nl<}(t) \, 
| N-1 , l-1, 0\rangle \, + \nonumber   \\
&-&  \beta \, \sqrt{{(l+1)(l+2)} \over {(2l+1)(2l+3)} }\, B_{Nl>}(t) \, 
| N-1 , l+1, -1 \rangle \, + \nonumber  \\
&+&  \left. \beta \, \sqrt{{l(l-1)} \over {(2l+1)(2l-1)} }\, B_{Nl<}(t) \, 
| N-1 , l-1, -1 \rangle   \,\right\}  \, ,
\\
| \psi_4 \rangle_t &=& \sum_{Nl} \lambda_{Nl} \, \left\{ \,
\beta \,{{l+1} \over {\sqrt{(2l+1)(2l+3)}} } \, B_{Nl>}(t) \, 
| N-1 , l+1, 0\rangle \,+  \right. \nonumber \\
&+&\beta \,{{l} \over {\sqrt{(2l+1)(2l-1}} } \, B_{Nl<}(t) \, 
| N-1 , l-1, 0\rangle \, +  \nonumber \\
&+&\alpha \, \sqrt{{(l+1)(l+2)} \over {(2l+1)(2l+3)} }\, B_{Nl>}(t) \, 
| N-1 , l+1, 1 \rangle \, + \nonumber \\
&-& \left. \alpha \, \sqrt{{l(l-1)} \over {(2l+1)(2l-1)} }\, B_{Nl<}(t) \, 
| N-1 , l-1, 1 \rangle   \,\right\} \, . 
\end{eqnarray}
We have used here shortened notation: {\em e.g.} 
$ A_{Nl>}(t)= A_{Nl j=l+1 \slash 2}(t) $ . 

 The evolution of the WP in the FW representation is by far more simple.
As previously, the initial form of the WP is given by (\ref{lin1}). Repeating
the procedure described above we get expressions for the WP at an arbitrary
 time $ t $. 
Note, that the WP consists of eigenstates with positive energy only, so the
small component -- equal to zero -- has been omitted for clarity reasons:
\begin{equation}
| \Psi_{FW+}( t)  \rangle= \left( \,
\matrix{ | \phi_1 \rangle_t \cr | \phi_2  \rangle_t }
 \, \right) \, .
\end{equation}
Parts with spin up and down are written as
\begin{eqnarray}
| \phi_1 \rangle_t &=& \sum_{Nl} \lambda_{Nl} \,
\Big\{ \,
\alpha \, \, a_{Nl}(t) \, | N,l,0 \rangle + \beta \, \, b_{Nl}(t) \, 
| N,l,-1 \rangle  \Big\}  \\
| \phi_2 \rangle_t &=& \sum_{Nl} \lambda_{Nl} \, \Big\{  \,
\beta \, \, a_{Nl}(t) \, | N,l,0 \rangle + \alpha \, \,  b_{Nl}(t) \, 
| N,l,1 \rangle \, \Big\} \, ,  \\
\nonumber
\end{eqnarray}
 where
 \begin{eqnarray}
a_{Nl}(t) &=&{{l+1} \over {2l+1}} \,   \exp(-i \omega_{Nl>} t)
+{ {l} \over {2l+1} } \,   \exp (-i \omega_{Nl<} t) \\
b_{Nl}(t) &=& {{ \sqrt{l(l+1)} } \over {2l+1}} \, \bigg( \,
\exp(-i \omega_{Nl>} t)  - \exp (-i \omega_{Nl<} t)
 \, \bigg)  \, .   \\ \nonumber
 \end{eqnarray}
 
\section{Spin averages for a single 
$|N,l,0\rangle \,[\,\alpha |\uparrow\rangle + \beta |\downarrow\rangle \,]$ 
state} 

Firstly, let us discuss the behavior of the spin vector defined by the 
average values of spin operators in the case of the state 
\begin{equation}
| \phi ( t=0) \rangle= |Nl0 \rangle \, 
\left( \,
\matrix{
\alpha \cr \beta \cr 0 \cr 0
} \,
\right) \, ,
\end{equation}
which is one of the component states of the linear WP (\ref{lin1}).
In the FW representation we easily obtain:
\begin{eqnarray}\label{snglsp}
 \langle  \sigma_x \rangle_t &=&
 2 \alpha \beta \, \bigg\{ \,
{{(l+1)^2+l^2} \over {(2l+1)^2}} +{ {2l(l+1)} \over {(2l+1)^2}} \,
\cos   (\omega_{Nl}\, t)   
 \, \bigg\}  \nonumber \\ 
\langle  \sigma_y \rangle_t &=& 0  \\
 \langle  \sigma_z \rangle_t &=&
  \big( \alpha^2 -  \beta^2 \big)  \,
\bigg\{ \,  {1 \over{(2l+1)^2 }} + { {4l(l+1)} \over {(2l+1)^2}} \,
\cos  (\omega_{Nl} \, t)    
\, \bigg\} \, , \nonumber 
\end{eqnarray}
where $ \omega_{Nl}=\omega_{Nl>} - \omega_{Nl<} $.
Hence, the vector of spin averages consists of two parts.
Both of them lie in the $ xOz $ plane. One is a constant vector
which inclination angle $\theta_1$
with respect to the $ Oz $ axis is given by:
\begin{equation}
\tan \theta_1 = {{2 \alpha \beta } \over { \alpha^2 - \beta^2}} \, 
[ \, (l+1)^2 +l^2 \, ] \, .
\end{equation}
The second part of the spin vector oscillates with the period
$ 2 \pi \slash \omega_{Nl} $. 
The average values of angular momentum behave in a similar way:
\begin{eqnarray}
\langle L_x \rangle_t &=& \alpha \beta \,
{{2l(l+1)} \over {(2l+1)^2}} \, [ \,1-\cos (\omega_{Nl}t) \, ] \nonumber \\
\langle L_y \rangle_t &=& 0 \\ 
\langle L_z \rangle_t &=& ( \alpha^2 - \beta^2 ) \,
{{2l(l+1)} \over {(2l+1)^2}} \, [ \,1-\cos (\omega_{Nl}t) \, ] \, ,\nonumber
\end{eqnarray}
It is easy to check that the mean value of the total angular momentum 
$ \langle {\mathbf J}\rangle =
\langle {\mathbf L} + 1 \slash 2 \mbox{ \boldmath $\sigma$}\rangle$
is conserved during evolution:
\begin{eqnarray}
\langle J_x \rangle_t &=& \alpha \beta  \nonumber\\
\langle J_y \rangle_t &=& 0 \\
\langle J_z \rangle_t &=& (\alpha^2 - \beta^2) \slash 2  \nonumber
\end{eqnarray}
and that the length of this vector is
$\sqrt{\sum_{i=1}^3 \langle J_i\rangle_t^2} = 
\sqrt{\frac{1}{4}(\alpha^2+\beta^2)^2}
= \frac{1}{2}$, because for linear WP ${\mathbf L}=0$.

\section{Autocorrelation function and spin averages}

An autocorrelation function at given time instant $ t $ is usualy defined 
as the projection of the WP $ | \Psi(t) \rangle $ onto the initial WP,
that is
\begin{equation}
A(t)=\langle \Psi (0) | \Psi (t) \rangle \, .
\end{equation}
Hence, the autocorrelation function (or rather square of its modulus,
which is a real quantity) is a widely used tool for illustrating
recurrences in WP's behavior. As $ 0 \le | A(t)|^2 \le 1 $,
the value of $ |A(t)|^2 $ indicates the degree of restoration of the 
initial shape and position of the WP. 
For our purposes we use a slightly modified (due to
relativistic regime) definition of autocorrelation function:
\begin{equation}
A(t)= \int \Psi^{\dag}(\vec r ,0) \Psi(\vec r, t) \, d^3 r \, .
\end{equation}
This leads us to the following expression in the Dirac representation
\begin{eqnarray}
A(t) &=& \sum_{Nl} \, |\lambda_{Nl}|^2 \, \left\{ \,
{{l+1} \over {2l+1}} \, \left[
\, \cos ( \,\omega_{Nl>} t\,) - i {{\omega_0} \over {\omega_{Nl>}}} \,
\sin ( \,\omega_{Nl>} t\,  ) \right] \right. + \nonumber \\
& & \hspace{1.8cm}+ \left.   
 {{l} \over {2l+1}} \, \left[
\, \cos ( \,\omega_{Nl<} t\,) - i {{\omega_0} \over {\omega_{Nl<}}} \,
\sin ( \,\omega_{Nl<} t\,  ) \right] \, 
\right\} \, , \\
\nonumber
\end{eqnarray}
 and in the FW representation
\begin{equation}
A_{FW} (t) = \sum_{Nl}  \, |\lambda_{Nl}|^2 \,
\left[ \,
{{l+1} \over {2l+1}} \, \exp( \, - i \omega_{Nl>}  \, t \, ) +
{{l  } \over {2l+1}} \, \exp( \, - i \omega_{Nl<}  \, t \, ) 
\, \right] \, '
\end{equation}
as well.
The calculation of spin averages for the full WP in the Dirac representation 
is straightforward but rather tedious:
\begin{eqnarray}
\langle \sigma_x \rangle_t &=& \alpha \beta \,\sum_{Nl}\, \left\{
|\lambda_{Nl}|^2 \, \left[
{ 1 \over 2} \left( {{l+1} \over {2l+1} }\right)^2 \,
\left( 1+{{\omega_0^2} \over{\omega_{Nl>}^2}}+
\left( 1-{{\omega_0^2} \over{\omega_{Nl>}^2}}
\right)
  \cos(2 \omega_{Nl>} t ) \right) + \right. \right. \nonumber \\
&+&  \hspace{3.0cm}
{ 1 \over 2} \left( {{l} \over {2l+1} }\right)^2 \,
\left( 1+{{\omega_0^2} \over{\omega_{Nl<}^2}}+
\left( 1-{{\omega_0^2} \over{\omega_{Nl<}^2}}
\right)
 \cos(2 \omega_{Nl<} t ) \right) + \nonumber  \\
&+& 
 {{l(l+1)} \over {(2l+1)^2} }  \,
\left( \left(
 1+{{\omega_0^2} \over{\omega_{Nl>} \omega_{Nl<} }} \, \right)
 \cos \left[ \, (\omega_{Nl>}-\omega_{Nl<}) \, t \, \right]
\right.+ \nonumber \\
&+& \hspace{1.95cm}
\left. \left(
 1-{{\omega_0^2} \over{\omega_{Nl>} \omega_{Nl<} }} \, \right)
 \cos \left[ \, (\omega_{Nl>}+\omega_{Nl<}) \, t \, \right]
 \right)+ \\
&-& 
{{(l+1)^2} \over {(2l+1)(2l+3)}} \,
\left(  1-{ {\omega_0^2} \over {\omega_{Nl>}^2}} 
\right) \, \big(  1- \cos (2 \omega_{Nl>} t)
\, \big)  +\nonumber \\ 
&-& \left. 
{{l^2} \over {(2l+1)(2l-1)}} \,
\left(  1-{ {\omega_0^2} \over {\omega_{Nl<}^2}} 
\right) \, \big(  1- \cos (2 \omega_{Nl<} t)
\, \big) \right] +\nonumber \\ 
&-& 
  {\rm Im}(\lambda_{Nl}^* \lambda_{Nl+2}) \,
{ {(l+1)(l+2)} \over {(2l+3) \sqrt{(2l+1)(2l+5)}}} \, \times \nonumber \\
& & \left. \hspace{3.0cm} \times \, 
{{ \sqrt{(\omega_{Nl>}^2 -\omega_0^2)(\omega_{Nl+2<}^2 -\omega_0^2) }  } 
\over { \omega_{Nl>} \omega_{Nl+2<} }} \,
\sin ( \omega_{Nl>} t) \sin ( \omega_{Nl+2<} t) \, \right\} \nonumber \\
\nonumber
\end{eqnarray}
\begin{eqnarray}
\langle \sigma_y \rangle_t &=& 0 \\ \nonumber
\end{eqnarray}
\begin{eqnarray}
\langle \sigma_z \rangle_t &=&
( \alpha^2 + \beta^2) \, \sum_{Nl} 
\left\{ \,
|\lambda_{Nl}|^2 \left[ 
{{l+1} \over {2(2l+1)^2}} \,
\left( 1+{{\omega_0^2} \over{\omega_{Nl>}^2}}+
\left( 1-{{\omega_0^2} \over{\omega_{Nl>}^2}}
\right)
 \cos(2 \omega_{Nl>} t ) \right) + \right. \right. \nonumber \\
 &-&  \hspace{2.3cm}
 {{l} \over {2(2l+1)^2}} \,
\left( 1+{{\omega_0^2} \over{\omega_{Nl<}^2}}+
\left( 1-{{\omega_0^2} \over{\omega_{Nl<}^2}}
\right)
 \cos(2 \omega_{Nl<} t ) \right) + \nonumber  \\
&+& 
 {{2l(l+1)} \over {(2l+1)^2} }  \,
\left( \left(
 1+{{\omega_0^2} \over{\omega_{Nl>} \omega_{Nl<} }} \, \right)
 \cos \left[ \, (\omega_{Nl>}-\omega_{Nl<}) \, t \, \right]
\right.+ \nonumber \\
&+& \hspace{2.0cm}
\left. \left(
 1-{{\omega_0^2} \over{\omega_{Nl>} \omega_{Nl<} }} \, \right)
 \cos \left[ \, (\omega_{Nl>}+\omega_{Nl<}) \, t \, \right]
 \right)+  
\nonumber   \\
&-& 
{{l} \over {2(2l+1)(2l+3)}} \,
\left(  1-{ {\omega_0^2} \over {\omega_{Nl>}^2}} 
\right) \, \big(  1- \cos (2 \omega_{Nl>} t)
\, \big)  
+  \\ 
&+& \left. 
{{1} \over {2(2l+1)(2l+3)}} \,
\left(  1-{ {\omega_0^2} \over {\omega_{Nl<}^2}} 
\right) \, \big(  1- \cos (2 \omega_{Nl<} t)
\, \big)
\right] 
+\nonumber \\ 
&+& 
  {\rm Re}(\lambda_{Nl}^* \lambda_{Nl+2}) \,
{ {4(l+1)(l+2)} \over {(2l+3) \sqrt{(2l+1)(2l+5)}}} \, \times \nonumber  \\
& & \left. \hspace{3.0cm} \times \, 
{{ \sqrt{(\omega_{Nl>}^2 -\omega_0^2)(\omega_{Nl+2<}^2 -\omega_0^2) }  } 
\over { \omega_{Nl>} \omega_{Nl+2<} }} \,
\sin ( \omega_{Nl>} t) \sin ( \omega_{Nl+2<} t) \, \right\} \nonumber \\
\nonumber
\end{eqnarray}
It is easy to see that terms containing $ \cos [ \,( \omega_{Nl>} - 
\omega_{Nl<}) t  \,] $ and constant terms are dominant ones.

Corresponding expressions are substantially simplified in the FW representation:
\begin{eqnarray}
 \langle  \sigma_x \rangle_t &=&
 2 \alpha \beta \, \sum_{Nl} | \lambda_{Nl}|^2  \, \bigg\{ \,
{{(l+1)^2+l^2} \over {(2l+1)^2}} +{ {2l(l+1)} \over {(2l+1)^2}} \,
\cos   \omega_{Nl}\, t   
 \, \bigg\}   \\ 
\langle  \sigma_y \rangle_t &=& 0  \\
 \langle  \sigma_z \rangle_t &=&
  \big( \alpha^2 -  \beta^2 \big)  \, \sum_{Nl} | \lambda_{Nl}|^2  \, 
\bigg\{ \,  {1 \over{(2l+1)^2 }} + { {4l(l+1)} \over {(2l+1)^2}} \,
\cos  \omega_{Nl} \, t    
\, \bigg\} \, . 
\end{eqnarray}
These expressions are just sums of (\ref{snglsp}) taken with weight 
coefficients $ | \lambda_{Nl}|^2 $.
 
In Fig.~1 
the evolution of average values of spin operators is presented
for the circular WP with the spin directed initially along 
$Ox$ axis ({\em i.e.}\  $\alpha =\beta = 1/\sqrt{2}$). 
Conventionally, the circular orbit lies in $xOy$ plane. 
For $r=0.0001$ (left part) the behavior of $\langle \sigma_i\rangle$
resembles that obtained in the nonrelativistic
description (figs.~1 and 2 of\cite{arv1}). 
A rapid collapse combined with a precession of the
spin vector (defined by average values of spin operators) is observed. 
At time instants $t_n =  (n \slash 2)T$, 
where $ T= 2 \pi \slash \omega $ is a classical period of the 3dHO,
there are spin revivals. Due to the quasiperiodicity of the evolution 
in relativistic case (nonlinear spectrum) it is hard to say if
the spin restores with an opposite sign for $n$ odd, as it happens in 
nonrelativistic case. For the case $r=0.0001$ (weakly relativistic)
the effect of {\em Zitterbewegung} is only slightly manifesting in 
$\langle\sigma_z\rangle$. 
This phenomenon is much better pronounced in the strongly relativistic case 
with $r=0.5$ (central picture). In this case
all components oscillate more rapidly due to the interference among states with
negative and positive energies. The {\em Zitterbewegung} is removed in the FW
representation (right part). One can notice the fast decrease in the amplitudes 
of revivals for bigger value of~$r$. 

Fig.~2 
shows the short time evolution of spin averages for the linear WP
in Dirac representation
(compare these results with nonrelativistic case in fig.~4 of \cite{arv2}). 
Note, that there is no precession, one observe spin collapses and revivals only.
The transition from the pure state of the spin in a well-defined direction 
$\mathbf n$ to the mixed state is especially clear in the 
case of the spin lying initially in the
$xOy$ plane $(\theta_{\sigma} = \pi \slash 2)$. There is a noticeable
{\em Zitterbewegung} for the larger value of $r$ -- once again it disappears 
in the FW representation. The amplitude of this trembling motion 
changes with increasing 
angle between $Oz$ axis and the initial spin direction, with the maximum for
$ \theta_{\sigma} = \pi \slash 2 $.

Relativistic effects, like the 
{\em Zitterbewegung} and an increase of the revival time with respect to 
the HO period $T$ grow also with
increasing $z_0$ 
and consequently with the mean energy of the wave packet.
It is well seen in Fig.~3 that for small $z_0$ (small WP energy) the motion 
is almost periodic with time scale very close to the HO period, while
for higher energies deviations from the HO period 
and imperfections of revivals become substantial. 
 
\section{Long time behavior} 

We found the existence of fractional revivals during the long time evolution
of the considered system. Using the prescription given by Averbukh and
Perelman \cite{ave} we have estimated the revival time (of the second order) as
$T_{\rm rev}\approx T/r $ for small values of $r$. 
According to Averbukh and Perelman's scenario
the first full revival is expected at time $(1/2) T_{\rm rev}$
preceded by a series of fractional revivals. 
Fig.~4 
illustrates such revivals for the circular WP as well as for the linear one. 
For circular WP the standard fractional revivals at $t/T=1/8, 1/4, 1/3, 1/2$
are well pronounced. For linear WP, due to different topology and the 
existence of boundary conditions \cite{pict,aron} we see spin collapses 
at $t/T= 1/8$ and $1/4$ and fractional revivals only at $t/T= 1/3$ and $1/2$.


 \section{Probability densities} 
 
 Initial stages of the evolution of the highly relativistic $(r=0.5)$ circular
WP are presented in the Fig.~5 in spherical coordinates $\vartheta$ and 
$\varphi$. The probability density on the sphere with the radius corresponding
to the radius of the classical orbit is displayed. The case presented in the 
figure corresponds to the initial direction of the spin parallel to $Ox$ axis.
In both representations a large part of the WP does not move at all
because angular velocities originated from orbital and spin motion cancel in DO
\cite{do}.
There are also two 
smaller subpackets which rotate in opposite directions with the same 
absolute value of the angular velocity. 
These subpackets don't spread during their evolution and move close to
the classical trajectory ($|{\mathbf r}|=x_0$ and $\vartheta= \pi
\slash 2$). The smaller one consists mainly of states with negative energy
 eigenvalues. This part of the WP disappears (the corresponding part of
 the probability density is shifted to the bigger one) in the FW representation.

 Figures~6 and 7 display the evolution of linear wave packets
 starting from the center of the coordinate system 
 and from a turning point, respectively.  The initial Gaussian WP transforms
 into spreading rings (case of the spin along $Oz$ axis, {\em i.e.} 
 along the orbit) that move along the 
 classical trajectory. The details of this type of motion have been 
 described (for nonrelativistic case) in \cite{pict,ber,ber1}.
 Particularly interesting is a kind of squeezing of the WP starting
 from the turning point when it reaches the center of the potential
 at $t=(1/4)T$.
For other spin directions rings become crescents (see Fig.~8)
 because cylindrical symmetry is absent in the initial conditions. 
 For longer evolution (not shown) one sees that 
 the linear WP starting from one turning point does not reach the other. 
 Instead,  the WP is squeezed near the center of the coordinate system and
 turned back. This unusual behavior is caused by the extremely 
 strong spin-orbit coupling in the DO. 
 
 As in the previous case of the circular WP one may observe, 
 if the value of $r$ is big enough, two subpackets travelling in opposite 
 directions.  One of them is removed in the FW representation.
 This behavior is well illustrated in Fig.~9.

\pagebreak[4]

 \begin{figure} \label{fig1}
\caption{Time evolution of spin components average values 
for the circular WP with $ N=20 $. Results for two different values of
$ r $ (0.0001 and 0.5) are presented.}
 \end{figure}\vspace{-2mm} 
 \begin{figure} \label{fig2}
\caption{Short time evolution of  the spin averages for 
 the linear WP in Dirac representation and its dependnce
 on the initial spin direction. Two cases of $r$ values are
 presented ($r=0.0001$ and $r=0.01$), $ \langle \sigma_n \rangle $ is
 an average value of the spin projection on the initial spin direction.}
 \end{figure}\vspace{-2mm}
 \begin{figure}\label{fig3}
\caption{ Comparison of the spin averages short time evolution for different
initial positions of the linear WP with r=0.01. Results for Dirac (left) and FW
(right) representations are presented. } 
 \end{figure}\vspace{-2mm}
 \begin{figure}[b] \label{fig4}
\caption{ Spin revivals in the long time evolution of the circular (top)
and the linear (bottom) WP in a weakly relativistic case. } 
 \end{figure}\vspace{-2mm}
 \begin{figure}\label{fig5}
\caption{ Initial stages of the evolution of the circular WP ($N=20, r=0.5$)  
 in Dirac and FW representations. Angular coordinates $\vartheta$  and 
 $\varphi$ are used to present the cross-sections of $\Psi^{\dag} \Psi$
 with the sphere $|{\mathbf r}|=z_0$.} 
 \end{figure}\vspace{-2mm}
 \begin{figure}\label{fig6}
\caption{ Evolution of the linear WP starting from the center of the coordinate
system ($z_0=0$). $ \Psi^{\dag} \Psi $  in the $xOz$ plane (containing the
classical trajectory) is presented. Remember that there is a cylindrical
symmetry with respect to $Oz$ axis.  } 
 \end{figure}\vspace{-2mm}
 \begin{figure}\label{fig7}
\caption{  The same as in the Fig. 6. but for the WP starting from the turning
point $z_0$ ($p_0=0$).  } 
 \end{figure}\vspace{-2mm}
 \begin{figure}\label{fig8}
\caption{   Cross-sections of the linear WP with the planes 
perpendicular to the classical trajectory. The axial symmetry is lost
if the initial spin direction is not along the $Oz$ axis.} 
 \end{figure}\vspace{-2mm}
 \begin{figure}\label{fig9}
\caption{Comparison of the evolution of the strongly relativistic 
linear WP (r=0.5) in Dirac and FW representations.  } 
 \end{figure}\vspace{-2mm}

\vfill

\end{document}